\documentclass[12pt]{iopart}

\usepackage{epsf}

\begin{document}
\jl{4}

\title[TeV $\gamma$-ray observations at large zenith angles]
{Effectiveness of TeV $\gamma$-ray observations at large zenith angles 
with a stereoscopic system of imaging atmospheric \v{C}erenkov telescopes}

\author{A.~Konopelko, F.~Aharonian, M.~Hemberger, W.~Hofmann, J.~Kettler, 
G.~P\"{u}hlhofer, H.J.~V\"{o}lk}

\address
{Max-Planck-Institut f\"ur Kernphysik, D-69029 Heidelberg, Germany}

\begin{abstract}
{The sensitivity of imaging atmospheric \v{C}erenkov telescopes (IACTs)  
 in TeV $\gamma$-ray observations reachs its maximum at small 
 zenith angles ($\Theta \leq 30^\circ$) which provide the minimum
 attainable energy threshold of an instrument. However, for a specific 
 telescope site a number of $\gamma$-ray sources, or source candidates, 
 can only be observed at much larger zenith angles ($\Theta \leq 60^\circ$). 
 Moreover the observations at large zenith angles allow to extend 
 the observation time window for any object seen at small 
 zenith angles, as well as to enlarge the dynamic energy range of an 
 instrument towards the highest observable energies of $\gamma$-rays. Based 
 on Monte Carlo simulations we present here the results on the   
 sensitivity of a stereoscopic system of 5 IACTs in observations at large 
 zenith angles. We point out some important parameters of the telescope 
 design which could substantially improve the efficiency of such observations 
 with forthcoming IACT arrays like CANGAROO~III, HESS and VERITAS.}
\end{abstract}

\maketitle

\section{Introduction}

Ground-based very high energy (VHE) $\gamma$-ray astronomy explores the 
energy range from 200~GeV up to 20~TeV. Recent exciting detections and 
observations of a number of $\gamma$-ray sources have demonstrated the very 
high sensitivity of the imaging atmospheric \v{C}erenkov technique (Weekes~et~al. 
1997; Ong, 1998). The performance of this technique appears to be strongly 
dependent on the zenith angle range covered while tracking an object. The 
accessible zenith angle range is simply determined by the latitude of the 
detector and the celestial coordinates of a particular object. The $\gamma$-ray 
observations are likely to be made up to zenith angles $30^\circ$ in order to 
detect high quality two-dimensional angular {{\it images} of \v{C}erenkov light 
from air showers. Extension of observations to larger zenith angles (up to 
$60^\circ$) might substantially widen the observational time window for a number 
of objects. Apparently that is very important for multiwavelength campaigns 
involving ground-based and satellite-born instruments in simultaneous observations 
of variable $\gamma$-ray sources, e.g. BL Lac objects. In addition observations at 
large zenith angles favour the detection of the high energy $\gamma$-rays, 
$\rm E \geq 10\,TeV$, which, at present, can be registered only using the 
ground-based \v{C}erenkov technique, and set important constrains on $\gamma$-ray 
emission mechanisms. 

Sommers \& Elbert (1987) first noticed that the effective detection area for 
$\gamma$-ray air showers dramatically increases in observations at large zenith 
angles using \v{C}erenkov detectors. Expedient use of large zenith angle observations 
using imaging \v{C}erenkov technique was suggested in the proposal of the HEGRA IACT 
array (Aharonian~et~al., 1989). Hillas \& Patterson (1991) made first simulations of 
\v{C}erenkov light images from air showers with an inclination of $60^\circ$. The 
Whipple group has developed the analysis technique for a single stand-alone 10 m telescope 
(see e.g. Krennrich et al. 1999) and tested it by observations of Mkn~421 and Mkn~501. 
Large zenith angle observations have been used with the 3.8~m \v{C}erenkov telescope 
operated by the CANGAROO group (Tanimori et al., 1998) to measure the energy spectrum of 
the Crab Nebula up to $\simeq 50$~TeV. Here we present results from Monte Carlo simulations 
of inclined air showers (zenith angle of $60^\circ$) for a {\it stereoscopic system} of 
five imaging air \v{C}erenkov telescopes (IACTs). We discuss the major change in topology 
of \v{C}erenkov light emission from air showers at large inclinations, which determines 
the detection rates as well as the ability to classify images of $\gamma$-rays. 
We have compared the sensitivity of the IACT system at small ($20^\circ$) (SZA) and 
large ($60^\circ$) (LZA) zenith angles. Current observations with the HEGRA IACT system 
generally confirm the Monte Carlo predictions (these results will be published elsewhere). 

\section
{Air shower simulations}

The ALTAI Monte Carlo code (Konopelko et al. 1996; 1999) was used to generate 
$\gamma$-ray and cosmic ray induced air showers. This code includes a detailed model of 
Rayleigh (molecular) scattering, aerosol (Mie) scattering and ozone absorption of \v{C}erenkov 
light in the atmosphere (Elterman, 1968; Driscoll, Vaughan, 1978). This model has been checked 
against the available experimental data for La Palma, Canary Islands (Hemberger, 1998). The 
calculations have been done with allowance for the effect of the geomagnetic field. By additional 
test simulations we found that the effect of multiple-scattering of \v{C}erenkov light photons in 
the atmosphere is negligible because of a low aerosol content above the observation level (2.2 km 
above sea level) as well as due to the specific shape of the aerosol scattering-phase function 
(Driscoll, Vaughan, 1978) for SZA as well as for LZA.

Simulations were performed for the setup of five IACTs of the HEGRA stereoscopic system 
(Aharonian~et~al. 1999). Each of five telescopes consists of a 8.5 $\rm m^2$ reflector 
focussing onto a photomultiplier tube camera. The number of photomultipliers in the camera 
was 271, arranged in a hexagonal matrix covering a field of view with a diameter  
of $4.3^\circ$. A telescope camera was triggered when the signal in two of the 271 
photomultiplier tubes exceeded a threshold of 8 photoelectrons, and the event was saved  
when at least two telescopes were triggered by \v{C}erenkov light from air shower. 
The overall efficiency of the photon-to-photoelectron conversion was about $0.1$. 

\section
{Lateral distribution of \v{C}erenkov light photons}

The longitudinal profile of the number of secondary electrons in air showers of a fixed primary 
energy, $\rm E$, remains almost the same at different shower inclinations when the 
distance along the shower axis is measured in $\rm gr/cm^2$. The depth of shower maximum can 
be calculated as $\rm T_{max} = X_0 \cdot ln(E/E_c)$ ($\rm X_0$ and $\rm E_c$ are the 
radiation length and critical energy in air, respectively) (e.g., Nishimura, 1967). 
Thus a 3 TeV $\gamma$-ray air shower at zenith has, on average, a maximum in electron 
number at a depth of $\rm T_{max} \simeq 400\, gr/cm^2$, which corresponds to 
$\rm H_{max} \simeq 7.5$~km above the sea.  
For large inclinations the total atmospheric optical depth increases substantially and the shower 
develops entirely in the upper layers of the atmosphere. For air showers with large inclinations, 
the height of shower maximum is far above the observation level, and the geometrical distance 
from the shower maximum to the observer is correspondingly very large. Thus a $\gamma$-ray-induced 
air shower of 3 TeV with $60^\circ$ inclination on average has an electron maximum at 
$\rm H_{max}\simeq 12$~km above the observation level. The corresponding geometrical distance 
from the shower maximum to observer, $\rm L_{max}$, for such a shower is about $24$~km.   
These geometrical factors determine the distribution of \v{C}erenkov light in the plane perpendicular 
to the shower core. Air showers at the zenith give a high \v{C}erenkov photon density in the region close 
to the shower core ($\rm R \leq \, 100 \, m$) (see Figure~1) because of high photon emission 
from the low energy electrons ($\rm E_e \leq 1\,GeV$) which deeply penetrate into the atmosphere 
and suffer multiple Coulomb scattering. In air showers with large inclinations these electrons 
contribute over a much wider range of distances from the shower core because of the larger geometrical 
distance to the shower maximum, and as a result the corresponding mean 
photon density is substantially lower. In addition, the absorption of \v{C}erenkov light photons in 
the atmosphere increases for air showers with large inclinations due to the large optical depth on the 
way from the point of their emission to the detector. 

The characteristic hump in the lateral distribution of 
\v{C}erenkov light photons, caused by the emission of energetic electrons ($\rm E_e \geq 1\,GeV$) 
around the shower maximum, is shifted to larger distances from the shower axis. One can estimate 
the position of the hump using the expression $\rm R_0 \sim L_{max} \cdot tg \theta_c$, where 
$\rm \theta_c = \theta_c(H_{max})$ is the \v{C}erenkov light emission angle at the corresponding 
shower maximum height, $\rm H_{max}$. Thus for a 3~TeV $\gamma$-ray shower at the zenith and for 
$60^\circ$ inclination the offset of the hump is 90 and 240~m, respectively. These 
``toy model'' estimates are in good agreement with the simulations (see Figure~1).
The hump becomes even more prominent at $60^\circ$ inclination because of the reduced density of 
\v{C}erenkov light photons emitted by low energy electrons. 

\section 
{Collection areas and detection rates}

Despite the sophisticated trigger logics, the trigger condition for each telescope of a 
system ultimately relies on the size of the image (i.e., total number of ph.-e.) in the camera. Thus  
the trigger efficiency for each telescope, as well as for the entire system roughly reflects the lateral 
distribution of \v{C}erenkov light photons at the observation level. Finally, the effective collection 
area is governed by the lateral distribution of \v{C}erenkov light photons  
(for details see Aharonian~et~al., 1995). Results of calculations for a 5 IACT system are shown in 
Figure~2. A sharp increase in collection areas at low energies, caused by increasing trigger efficiency 
within the plateau of \v{C}erenkov light density ($\rm R\geq \,130\, m$), changes above $\rm S_\gamma 
\sim 5 \cdot 10^4 \, m^2$ to a logarithmic growth at the exponential tail of the \v{C}erenkov light 
lateral distribution (see Figure~1). At large inclinations low energy $\gamma$-rays ($\rm E \leq 1\, TeV$) 
cannot trigger the telecope system because of the very low average image size. At the same time the broad 
lateral distribution for high energy $\gamma$-rays ($\rm E \geq 3\,TeV$) provides large collection areas 
which could even substantially exceed the collection areas at small zenith angles (see Figure~2). Thus, for 
10 TeV the $\gamma$-ray collection area for LZA is larger by a factor of 3.5 than for SZA. 

\begin{table}
\caption{Detection rates of $\gamma$-ray and cosmic ray-induced air
  showers for inclination of $20^\circ$ and $60^\circ$ and  
  corresponding signal to noise ratio for all triggered events and 
  for the high energy events above 20~TeV.}
\begin{indented}
\item[]\begin{tabular}{@{}llll} \br 
           & Zenith angle:             & $20^\circ$       & $60^\circ$\\ \mr
           & $\rm R_\gamma,\, [hr^{-1}]$ & 100              & 43\\
All events & $\rm R_{CR},\,   [hr^{-1}]$ & $5.4 \cdot 10^4$ & $2.7 \cdot 10^4$\\ 
           & S/N, $\sigma$                       & 0.3     & 0.18 \\ 
           & $\rm R_\gamma,\, [hr^{-1}]$ & 1.2              & 5\\
$\rm E \ge 20\, TeV$& $\rm R_{CR},\,   [hr^{-1}]$ & $8\cdot 10^2$    & $1.6\cdot 10^3$\\
           & S/N, $\sigma$              & 0.03    & 0.09 \\ \br
\end{tabular}
\end{indented}
\end{table}

\begin{table}
\caption{Effective energy threshold of $\gamma$-ray induced air showers at different 
         inclination angles.}
\begin{indented}
\item[]\begin{tabular}{@{}llllll} \br
$\Theta$, deg     & 0   & 20  & 30  & 45  & 60  \\ \mr 
$\rm E_{th}$, TeV & 0.5 & 0.7 & 0.9 & 1.8 & 5.0 \\ \br
\end{tabular}
\end{indented}
\end{table}

Assuming an energy spectrum of $\gamma$-rays, e.g., $\rm dJ_\gamma/dE \propto E^{-\alpha};\,
\alpha = 2.5$, and a certain flux normalization, $\rm J_\gamma(>1\,TeV) = 10^{-11}\, cm^{-2}s^{-1}$, 
one can calculate the detection rates of $\gamma$-ray-induced air showers  
\begin{equation}
R_\gamma (>E_o) = \int_{E_o} S_\gamma(E)\frac{dJ_\gamma}{dE}dE
\end{equation}
and cosmic ray showers  
\begin{equation}
R_{CR} (> {E_o}) =\int_{\Omega_o}d\Omega \int_{E_0} 
S_{CR}(E,\Omega)\frac{dJ_{CR}}{dE} dE
\end{equation}  
above the energy threshold of $\rm E_0$, where $\Omega_o$ is the solid angle of the isotropic 
cosmic ray air showers, and $S_\gamma$ and $S_{CR}$ are the collection areas for $\gamma$-rays and 
cosmic rays, respectively. $E$ is a reconstructed energy of $\gamma$-ray and 
cosmic ray induced air showers using the procedure tuned for the $\gamma$-rays 
(see Konopelko~et~al, 1999).  
The integral detection rates calculated for SZA and LZA  
are shown in Figure~3. Note that for $\gamma$-rays above 20 TeV the integral detection rate 
is about 4 times higher for LZA than for SZA, whereas the 
corresponding integral cosmic ray rate is higher by a factor of 2, only. Thus at  
the trigger level LZA observations reveal a substantial advantage in the detection 
rate of high energy $\gamma$-rays as well as a high signal to noise ratio, 
$\rm S/N = R_{\gamma}/(2 \cdot R_{CR})^{1/2}$ (see Table~1)
\footnote{In TeV $\gamma$-ray observations the significance of $\gamma$-ray signal is calculated as 
$\rm S/N = ON - OFF/ (ON+OFF)^{1/2}$ where ON is the number of events registered while tracking the 
source and OFF is the corresponding number of background events, $\rm N_{CR}$. ON - OFF gives the number of 
detected $\gamma$-ray showers, $\rm N_\gamma$. In the case when $\rm OFF>>ON-OFF$ one can 
modify the formulae as follows: $\rm N_\gamma/(2 \cdot N_{CR})^{1/2} = R_\gamma / (2\cdot R_{CR})^{1/2}\cdot 
\sqrt{t}$, where $t$ is the observational time and $\rm R_\gamma, R_{CR}$ are the rates of $\gamma$-rays and 
cosmic rays, respectively.}     
The signal to noise ratio for the high energy events (above 20~TeV) 
strongly depends on the energy spectrum index of $\gamma$-rays. In case of a flat 
energy spectrum ($\alpha \sim 2.0$) the advantage at LZA becomes even more 
prominent (see below). The effective energy threshold of the telescope system, defined as the energy 
at which $\gamma$-ray detection rate reaches its maximum for the differential energy spectrum 
$\rm dN_{\gamma}/dE \sim E^{-2.5}$, is $\sim 0.5$ TeV at zenith, and increases at larger 
inclinations (see Table~2). 

\section
{Orientation and shape of \v{C}erenkov light images}

The above mentioned features of the shower development at large inclinations determine the topology 
of \v{C}erenkov light images. We show in Figure~4 \v{C}erenkov light images calculated for 
3 TeV $\gamma$-ray-induced air shower with various inclinations. In general the images 
at large inclinations {\it (a)} contain less photons; {\it (b)} become smaller in size; {\it (c)} 
shrink to the camera center; and, {\it (d)} have a circular shape. Using our ``toy model'' 
considerations we may estimate the position of the image centroid (maximum of image intensity 
set by the emission from shower maximum) in a telescope focal plane as $\rm \theta_0 \sim 
1/tg(R_0/L_{max})$. At 100~m impact distance from the shower axis the centroid position is at 
$\simeq 1^\circ$ and $\simeq 0.23^\circ$, for air showers at the zenith and for $60^\circ$ inclination, 
respectively. The small angular size of \v{C}erenkov light images at large inclinations was noticed 
by Hillas \& Patterson (1990). Air showers with $60^\circ$ inclination are very far from the observer 
which is why the image angular size is getting small in both longitudinal and transverse directions. 
The lateral spread of electrons ($\rho_o$) in the maximum of multi-TeV $\gamma$-ray 
shower is $\sim 20$~m (see e.g., Hillas, 1996). Thus the angular size of the \v{C}erenkov light image 
along the minor image axis can be estimated as $w \rm \sim 1/tg(R_0/L_{max})-1/tg((R_0-\rho_0)/L_{max})$. 
It is of $\simeq 0.2^\circ$ for shower at zenith and $\simeq 0.06^\circ$ for the shower with $60^\circ$ 
inclination (see Figure~4). 
To measure the orientation and shape of these images one needs a relatively small pixel size 
(angular size of PMTs in camera) $\sim 0.1 \div 0.15^\circ$. Note also that light smearing by optical errors 
may significantly distort the angular shape of these images. The ratio of standard second 
moment parameters, {\it Width}/{\it Length}, shifts to larger values (see Figure~6) for showers at large 
inclinations. It shows that these images have a circular shape rather than an elongated 
elliptic shape. Consequently the image orientation is determined with larger uncertainties. 

\begin{table}
\caption{Acceptances of $\gamma$-ray showers after angular
  cut for two inclinations, $20^\circ$ and
  $60^\circ$. $\theta$ is the angular distance from the reconstructed
  to the actual $\gamma$-ray source position.} 
\begin{indented}
\item[]\begin{tabular}{@{}lll}\br
Zenith angle: & $20^\circ$ & $60^\circ$ \\ \mr 
$\theta \leq 0.3^\circ$    & 0.90 & 0.52 \\
$\theta \leq 0.55^\circ$   & 0.95 & 0.70 \\ \br
\end{tabular}
\end{indented}
\end{table}

\begin{table}
\caption{Acceptances of $\gamma$-ray and cosmic ray showers and corresponding 
  enhacement (Q-factor) after applying mean scaled Width cut. 
  Air showers were simulated at two inclinations, $\theta$ = $20^\circ$ and $60^\circ$.}
\begin{indented}
\item[]
\begin{tabular}{@{}lllll} \br
                        & Cut:                       & $\kappa_\gamma$            & $\kappa_{cr}$ & Q-factor\\ \mr 
$\theta = 20^\circ$     & $< \tilde w > \leq$ 1.0    & 0.53                       & 0.01          & 5.3     \\ 
                        & $< \tilde w > \leq$ 1.3    & 0.99                       & 0.15          & 2.6     \\ \mr
$\theta = 60^\circ$     & $< \tilde w > \leq$ 1.0    & 0.45                       & 0.05          & 2.0     \\
                        & $< \tilde w > \leq$ 1.3    & 0.90                       & 0.30          & 1.6   \\ \br 
\end{tabular}
\end{indented}
\end{table}

For the image orientation one can use the standard $Alpha$-parameter which defines the angle between 
the major axis of an elliptical image and the line connecting the position of the image maximum to the 
camera center (we assume that telescopes are looking directly onto the source of $\gamma$-rays). The 
distribution of $Alpha$-parameter for all triggered telescopes in a system observing the $\gamma$-rays 
at $20^\circ$ and $60^\circ$ inclination is shown in Figure~5. The distribution at large inclination is 
relatively broader as expected. This implies that $\gamma$-ray acceptance after applying the angular cut 
is less at large inclinations (see Table 3) and the corresponding enhancement factor is lower. Note that 
an imaging camera with small pixels ($\sim 0.1\div 0.15^\circ$) may substantially improve the angular 
resolution at large inclinations. 

In order to utilize simultaneously several \v{C}erenkov light images for an 
individual shower the so-called {\it mean scaled Width} parameter (Konopelko, 1995; Daum et al., 1996) 
can be effectively used for a system of IACTs. To compensate the dependence of the image shape on primary 
shower energy and distance from shower core to the telescope (impact parameter) the standard parameter 
{\it Width} (see Fegan, 1997) ($w^k$), calculated for each telescope, is scaled according to the Monte 
Carlo predicted values, $<w>^k_{ij}$, taken for the corresponding bin of reconstructed distance from 
the telescope to the shower core (i) and for the corresponding bin of image size (total number of 
photoelectrons in the image)(j). The {\it mean scaled Width} parameter is defined for each individual 
shower as follows
\begin{equation}
< \tilde w > = 1/N \sum_{k=1}^{N} w^k/<w>^k_{ij}
\end{equation} 
where $N$ is the number of triggered telescopes. The optimum cut on mean scaled Width is about 1.0, which 
gives a $\gamma$-ray acceptance of $\sim 50$\%. However, for a precise determination of $\gamma$-ray
spectra, a loose cut on mean scaled width ($<\tilde w> < 1.2$) has been so far used in data analysis 
(Aharonian et al, 1999) in order to maximize the $\gamma$-ray acceptance and to minimize systematic error related to cut efficiencies. In Table 4 we show the acceptances and efficiencies of a $\gamma$-ray 
classification using the {\it mean scaled Width} cut. The small angular size of both $\gamma$-ray and 
cosmic ray induced air showers prevents effective rejection using $<\tilde w >$-parameter for LZA. The 
resulting enhancement (Q-factor) does not exceed 2.0 whereas at small zenith angle it is more than 5.0.
We may conclude that the standard orientation and shape cuts, usually used for the IACT system analysis, 
allow less significant rejection of the cosmic rays.  In order to improve the cosmic ray rejection we 
introduced an additional parameter, {\it mean scaled Length}, $<\tilde l>$, defined by analogy with 
$<\tilde w>$. Two parameters, $<\tilde w>$ and $<\tilde l>$, can be used for calculating a Mahalanobis 
distance, MD (see Mahalanobis, 1963), in 2-dimensional space 
\begin{equation}
\rm MD = ((1-<\tilde w>)^2/\sigma^2_{<\tilde w>}+(1-<\tilde l>)^2/\sigma^2_{<\tilde l>})^{1/2}
\end{equation}
where $\sigma_{<\tilde w>}$ and $\sigma_{<\tilde l>}$ are standard deviations for the corresponding
distributions of $<\tilde w>$ and $<\tilde l>$. We found that the optimum value for the MD cut for LZA is 
1.5. This analysis improves the enhancement factor by $\simeq 30$\%.  
Note that multivariate analysis technique for a single \v{C}erenkov telescope has been discussed before by 
Aharonian~et~al. (1991); Hillas \& West (1991). In addition we have applied the standard algorithms of 
the impact distance and energy reconstruction (e.g., Konopelko et al., 1999) used for the system of 5 IACTs 
to events simulated at large inclinations. A summary of a system performance at LZA is presented in Table 5. 

\section 
{Sensitivity to $\gamma$-ray fluxes}

The sensitivity of the instrument can be characterized using the so-called {\it sigma-per-hour} parameter. 
It corresponds to the signal to noise ratio which one can achieve within one hour observations of 
$\gamma$-ray source assuming a certain DC $\gamma$-ray flux, e.g., 
$J_\gamma (\geq 1~TeV) = 10^{-11}$ $\rm [photons/cm^2 s]$ and the energy spectrum 
$dJ_\gamma/E \propto E^{-2.5}$. Thus the sensitivity of the 5 IACT system is about $6\sigma$ and 
$\sim 1 \sigma$ (loose cuts) at the small ($20^\circ$) and large ($60^\circ$) zenith angles. These 
sensitivities can be explained by the difference in the energy threshold and $\gamma$-ray rate for LZA and 
SZA. Thus, for a proper comparison of system performance one can use a ``sigma-per-hour'' for energy bin, 
$\sigma/hr/log(E)$. This parameter corresponds to the signal to noise ratio which one could expect detecting 
$\gamma$-rays in a fixed energy range. In present analysis we choose the corresponding energy bin as 
$\rm \Delta E_i = E_{i+1} - E_i, \, E_{i+1}/E_i = 1.38$ which roughly corresponds to the energy resolution 
at LZA. The number of $\gamma$-ray and cosmic ray induced showers detected within the energy bin 
$\Delta E_i$ can be calculated as 
\begin{equation}
\rm N_i^\gamma = \int_{E_i}^{E_{i+1}}(\frac{dR^\gamma}{dE})\cdot dE, \,
\rm N_i^{CR} = \int_{E_i}^{E_{i+1}}(\frac{dR^{CR}}{dE})\cdot dE,
\end{equation}
and the corresponding signal to noise ratio is
\begin{equation}
\rm S_i = N_i^\gamma /(2 \cdot N_i^{CR})^{1/2}.
\end{equation}
where $\rm (dR^\gamma/dE)$ and $\rm (dR^{CR}/dE)$ are the differential detection rates of $\gamma$-ray and 
cosmic ray air showers. 

The $\sigma/hr/log(E)$ as a function of primary $\gamma$-ray energy, $\rm S = S(E)$, is shown in Figure~7. 
Our calculations demonstrate that for a $\gamma$-ray source with a spectrum index of 2.5, LZA observations 
have an advantage at trigger level before imaging analysis. After applying the software analysis cuts, the 
sensitivity is almost the same at $\rm 10~TeV$ with a slight advantage for LZA for higher energies. For 
flat-spectrum $\gamma$-ray sources ($\alpha \sim 2.0$) the siganal-to-noise ratio is a factor of 3 higher 
at $\sim 20$~TeV for LZA compared with SZA. For the data shown in Figure~7 we have applied the {\it loose} 
analysis cuts. These cuts keep most of the $\gamma$-rays ($\rm \kappa_\gamma \simeq 0.9$), which is 
preferable for the spectrum studies. However these cuts give only a modest cosmic ray background rejection 
with the corresponding quality factor of $\simeq 12$ for SZA and $\simeq 5$ for the LZA. Note that in order 
to achive the maximum signal-to-noise ratio one can use the {\it tight} cuts ($\theta \leq 0.22^\circ$, 
$< \tilde w> < 1.0$) which provide a $\gamma$-ray acceptance of $\kappa_\gamma \simeq 0.4$ and 
corresponding Q-factors as high as $\simeq 35$ and $\sim 10$ for SZA and LZA, respectively.        

\begin{table}
\caption{Accuracy of determination of impact distance ($\rm \Delta R$), angular resolution 
($\delta \Theta$), and energy resolution ($\rm \Delta E/E$), for a system of 5 IACTs.} 
\begin{indented}
\item[]
\begin{tabular}{@{}lll} \br
Zenith angle: & $20^\circ$ & $60^\circ$ \\ \mr
$\rm \Delta R$,~m  & $\leq 10$  & $\leq 50$  \\ 
$\delta \Theta$,~degree & 0.1 & 0.3      \\
$\rm \Delta E/E$,~\% & 20 & $\leq 30$ \\ \br
\end{tabular} 
\end{indented}
\end{table}

\section
{Conclusions}

We have studied the efficiency of large zenith angle observations using the stereoscopic system of 5 
imaging atmospheric \v{C}erenkov telescopes by means of detailed Monte Carlo simulations. LZA observations 
give a very large effective collection area at large zenith angles but a modest ability for $\gamma$-ray 
classification. The results of simulations show that for a $\gamma$-ray source with a relatively steep 
energy spectrum ($\alpha \geq 2.5$) LZA observations provide almost the same sensitivity as normal SZA 
observations at the energy of $\gamma$-rays of about 10~TeV with some advantage at higher energies. For 
$\gamma$-ray sources with a flat energy spectrum ($\alpha_\gamma \sim 2.0$) LZA observations have a 
significantly higher sensitivity at energies $\rm \geq 10~TeV$.     

Observations at large zenith angles with the forthcoming arrays of imaging atmospheric \v{C}erenkov 
telescopes, like CANGAROO~III, HESS and VERITAS, with an energy threshold of 50-100~GeV, could provide an 
extension of dynamic energy range up to 20-50~TeV. Such observations would need a camera with small pixel 
size of $\sim 0.1\div 0.15^\circ$ and fine optics of the telescope reflector. 

\section{References}
\begin{harvard}

\item
Aharonian, F., et al,. 1989 {\it Proposal for Imaging Air \v{C}erenkov Telescopes 
in the HEGRA Particle Array} 

\item
Aharonian, F., et al., 1991 {\it NIM in Phys. Res.} {\bf A302} 522-528

\item
Aharonian, F., et al., 1995 {\it J. Phys. G: Nucl. Part. Phys.} {\bf 21} 419-428

\item
Aharonian, F., et al., 1999 {\it Astron. Astrophys.} {\bf 342} 69-86

\item
Daum, A., et al, 1998, {\it Astroparticle Physics}, Vol. 8, Nos. 1-2, p. 1

\item
Driscoll, W., Vaughan, W. 1978 {\it Handbook of optics} MCGRAW-HILL book company

\item
Elterman, L. 1968 {\it AFCRL-68-0153}, Bedford, Massachusetts

\item
Fegan, D. 1998, {\it J. Phys. G: Nucl. Part. Phys.} 23, 1013

\item
Hemberger, M. 1998 {\it PhD thesis}, Heidelberg

\item
Hillas, A.M., Patterson, J. 1990 {\it J. Phys. G: Nucl. Part. Phys.} {\bf 16} 1271-1281 

\item
Hillas, A.M., 1996 {\it Space Science Reviews}, vol. 75, N 1-2, 17 

\item
Hillas, A.M., West, A. 1991 {\it Proc. 22nd ICRC}, Dublin, 1, 472

\item
Konopelko, A., 1995, in Proc. 
{\it Towards a Major Atmospheric \v{C}erenkov Detector-IV},
Padova (ed. by M. Cresti), p. 373

\item
Konopelko, A., et al. (HEGRA Collaboration) 1996, {\it Astroparticle Physics}, 4, 199

\item
Konopelko, A., et al. (HEGRA Collaboration) 1999, {\it Astroparticle Physics}, 10, 275

\item
Krennrich, F., et al., 1999, {\it ApJ} {\bf 511}, 149

\item
Nishimura, J. 1967, {\it Nandbuch der Physik} XLVI/2, 1 

\item 
Ong, R. 1998, {\it Physics Reports}, 305, 3-4, 94

\item
Sommers, P., Elbert, J. 1987 {\it J. Phys. G: Nucl. Phys.} {\bf 13} 553-566 

\item
Tanimori T., et al., 1998 {\it ApJ} {\bf 492} L33-L36   

\end{harvard}

\begin{figure}
\begin{center}
\epsfxsize= 9 cm
\epsfbox{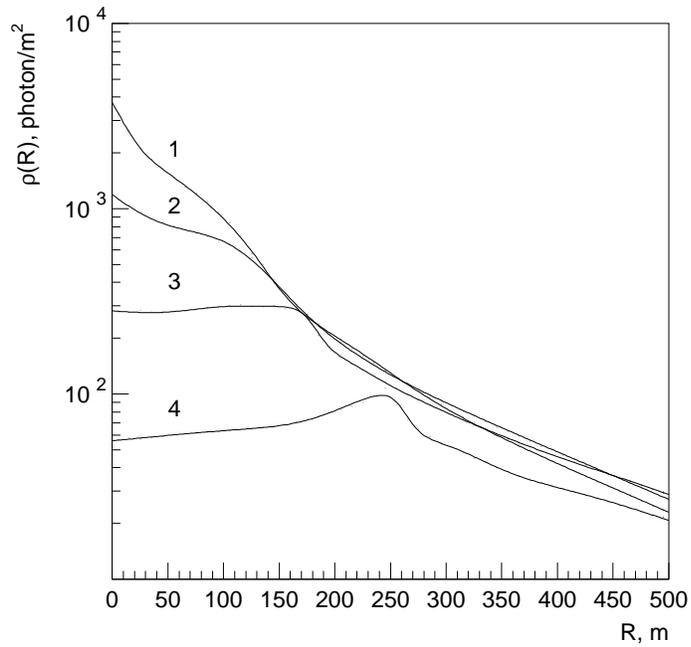}
\caption{\protect \small Lateral distribution of \v{C}erenkov light 
  density in 3 TeV $\gamma$-ray-induced air shower with inclination angle of  
  $0^\circ$(1); $30^\circ$(2); $45^\circ$(3) and $60^\circ$(4). The 
  observation level is about 2.2 km above the sea.}
\end{center}
\end{figure}

\begin{figure}
\begin{center}
\epsfxsize= 9 cm
\epsfbox{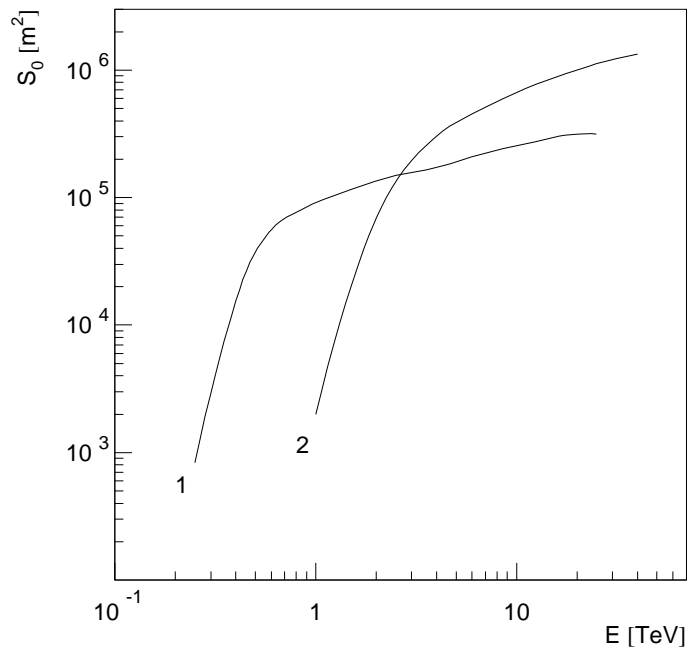}
\caption{\protect \small Detection area as a function of a primary shower energy  
  for $\gamma$-ray-induced air showers with inclination angle of $20^\circ$(1) 
  and $60^\circ$(2).}
\end{center}
\end{figure}

\begin{figure}
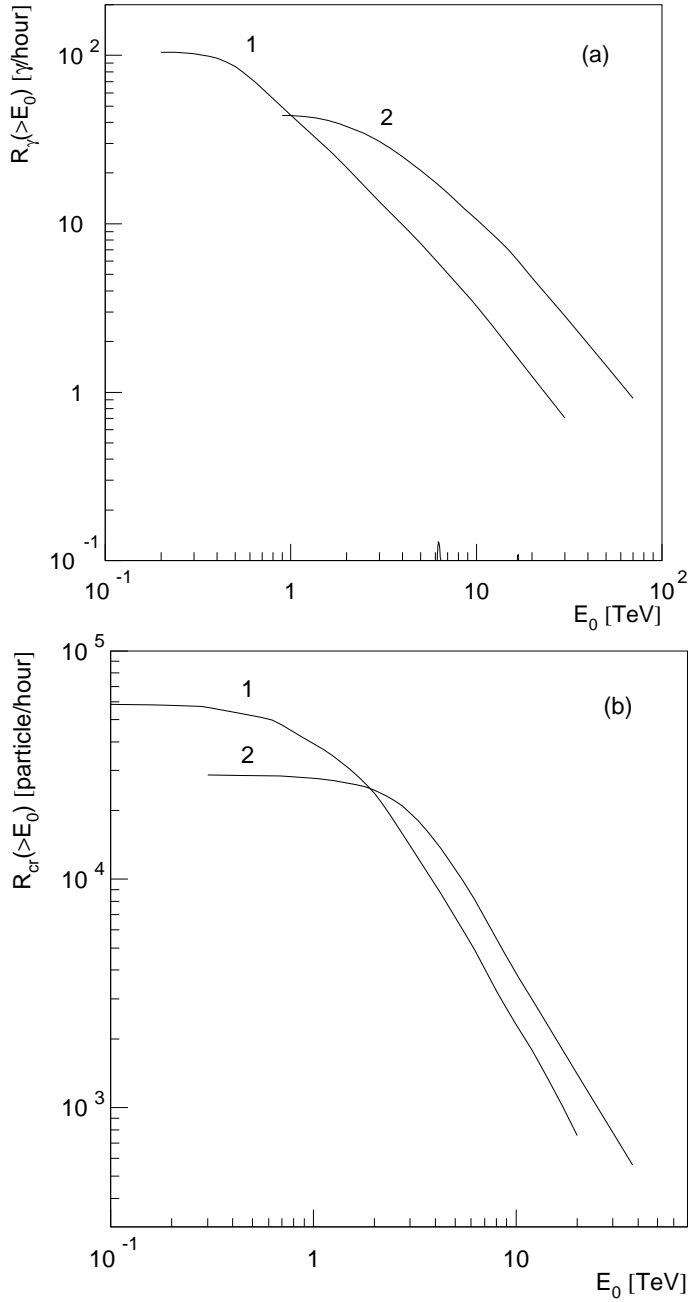

\begin{center}
\epsfxsize= 9 cm
\epsfbox{lza-fig3a.epsi}
\epsfxsize= 9 cm
\epsfbox{lza-fig3b.epsi}
\caption{\protect \small Integral detection rates of $\gamma$-ray (a) and 
  cosmic ray (b) air showers above the energy $\rm E_0$ for two inclination 
  angles, $20^\circ$(1) and $60^\circ$(2).  }
\end{center}
\end{figure}

\begin{figure}
\begin{center}
\epsfxsize= 7 cm
\epsfbox{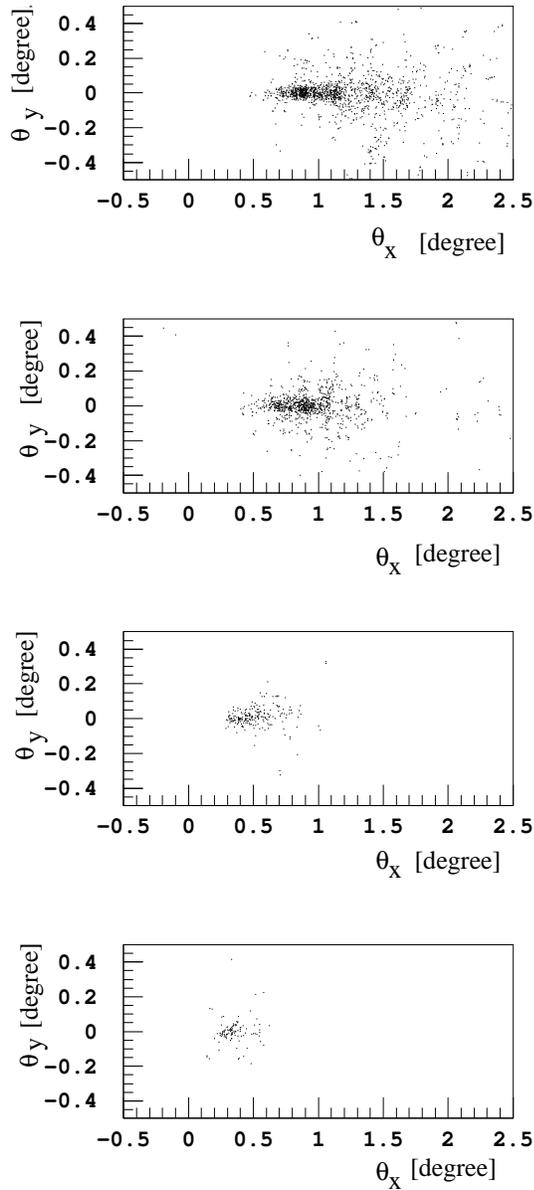}
\caption{\protect \small \v{C}erenkov light images of 3 TeV $\gamma$-ray air 
  shower simulated at different inclination angles, $0^\circ$, $30^\circ$, $45^\circ$ and 
  $60^\circ$ (from upper to bottom panel). The calculated images reproduce the two-dimensional 
  angular distributions ($\theta_x, \theta_y$) of \v{C}erenkov light photons hitting the telescope 
  reflector. Angle $\theta_x$ is measured respecting the axis which 
connects the telescope mirror and 
  the shower core. The impact distance from the shower core to the telescope is about 100~m.}
\end{center}
\end{figure}

\begin{figure}
\begin{center}
\epsfxsize= 9 cm
\epsfbox{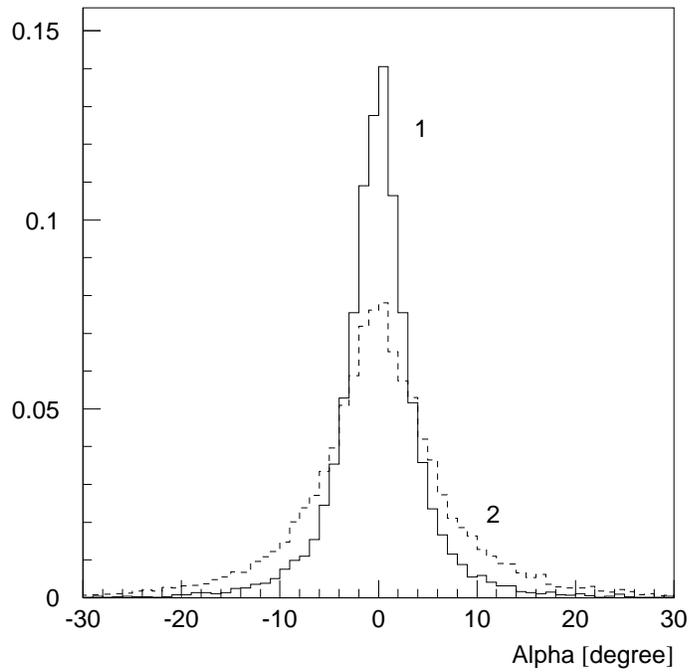}
\caption{\protect \small Distribution of image orientation parameter $Alpha$ for  
  $\gamma$-ray-induced air showers simulated at inclination angles of
  $20^\circ$(1) and $60^\circ$(2).}
\end{center}
\end{figure}

\begin{figure}
\begin{center}
\epsfxsize= 9 cm
\epsfbox{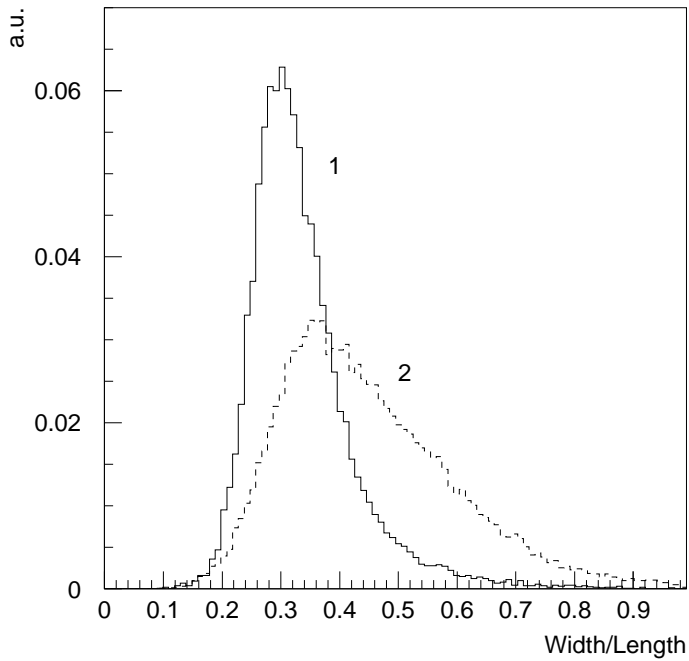}
\caption{\protect \small Distribution of {\it Width}/{\it Length} ratio 
  for $\gamma$-ray-induced air showers simulated for the inclination angle 
  of $20^\circ$(1) and $60^\circ$(2).}
\end{center}
\end{figure}

\begin{figure}
\begin{center}
\epsfxsize= 9 cm
\epsfbox{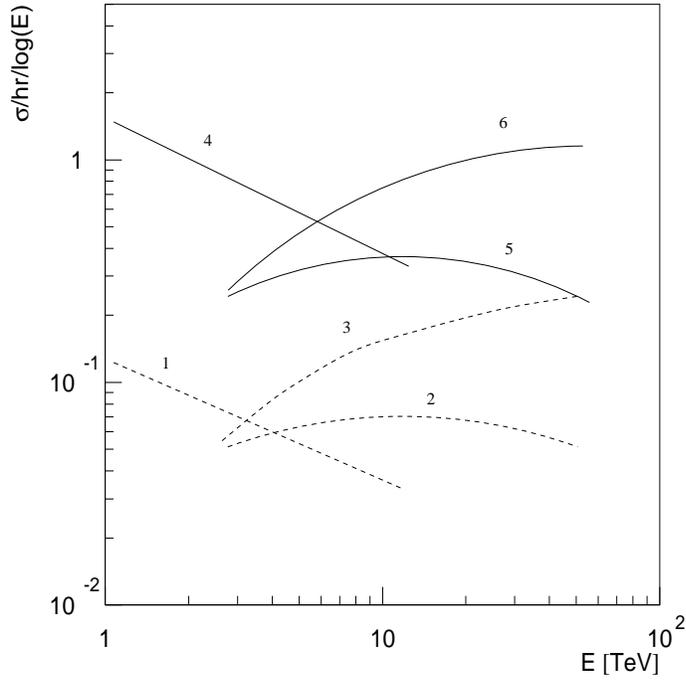}
\caption{\protect \small Expected signal-to-noise ratio (see eqn. 5) in 
  observations at small zenith angles ($20^\circ$) before (curve~1)
  and after (curve~4) {\it lose} analysis cuts, assuming the differential 
  energy spectrum $dJ_\gamma/dE \sim E^\alpha, \alpha = 2.5$ with normalization 
  $J_\gamma(>1\,TeV)=10^{-11}\, \rm cm^{-2}s^{-1}$ 
  as well as for the large zenith angles ($60^\circ$), before (2, $\alpha = 2.5$; 3,
  $\alpha=2.0$) and after software cuts (5, $\alpha = 2.5$; 6, $\alpha = 2.0$).}
\end{center}
\end{figure}

\end{document}